\documentclass[aps,prl,reprint,superscriptaddress,longbibliography,preprintnumbers]{revtex4-2}
\usepackage[T1]{fontenc}
\usepackage{amsmath,amssymb,amsthm}
\usepackage{mathtools,booktabs,longtable,tikz,microtype}
\usepackage{xcolor}
\definecolor{apsblue}{rgb}{0.18, 0.19, 0.57}
\usepackage[
    colorlinks=true,
    allcolors=apsblue,    
    breaklinks=true       
]{hyperref}
\usepackage{extarrows}

\definecolor{vTwoBlue}{HTML}{225F82}
\definecolor{vTwoOrange}{HTML}{C06522}
\definecolor{vTwoRim}{HTML}{9CAFB8}
\tikzset{
  vTwo bulk/.style={circle,fill=black,inner sep=1.3pt},
  vTwo fixed/.style={rectangle,draw=black,fill=white,inner sep=1.7pt},
  vTwo external/.style={draw=vTwoBlue,line width=.8pt},
  vTwo internal/.style={draw=black,line width=.95pt},
  vTwo boundary/.style={draw=vTwoRim,line width=.55pt}
}
\newcommand{\vTwoLeg}[3]{%
  \draw[vTwo external] #1--(#2:1);
  \fill[vTwoBlue] (#2:1) circle(1.2pt);
  \node[font=\scriptsize,inner sep=0pt] at (#2:1.27) {$#3$};}
\newcommand{\vTwoAuxLeg}[3]{%
  \draw[vTwoOrange,line width=.9pt] #1--(#2:1);
  \fill[vTwoOrange] (#2:1) circle(1.2pt);
  \node[font=\scriptsize,inner sep=0pt] at (#2:1.27) {$#3$};}
\newcommand{\vTwoDiagram}[2][.68]{%
  \vcenter{\hbox{\begin{tikzpicture}[scale=#1]
    \path[use as bounding box] (-1.42,-1.42) rectangle (1.42,1.42);
    \draw[vTwo boundary] (0,0) circle(1);
    #2
  \end{tikzpicture}}}}
\newcommand{\vTwoFixedRoot}{%
  \node[vTwo fixed] at (-.62,0) {};
  \node[font=\scriptsize,inner sep=0pt] at (-.62,-.35) {$X$};}

\newcommand{\beq}{\begin{equation}}
\newcommand{\eeq}{\end{equation}}

\begin{document}
\preprint{USTC-ICTS/PCFT-26-63}
\title{No Particle Production and Integrability in AdS$_2$}
\author{De-liang Zhong}
\affiliation{Institute of Theoretical Physics, Chinese Academy of Sciences, Beijing 100190, China}
\affiliation{School of Physical Sciences, University of Chinese Academy of Sciences, No. 19A Yuquan Road, Beijing 100049, China}
\author{Xinan Zhou}
\affiliation{Kavli Institute for Theoretical Sciences, University of
Chinese Academy of Sciences, Beijing 100190, China}
\affiliation{Peng Huanwu Center for Fundamental Theory, Hefei, Anhui 230026, China}
\date[]{} 
\begin{abstract}
    We propose an analogue of the no particle production condition in two-dimensional anti-de Sitter space, and demonstrate how this criterion uniquely reconstructs the curved-space Liouville and Toda theories via general-multiplicity holographic correlators. 
\end{abstract}
\maketitle

\section{Introduction}

In two dimensions, integrability links an infinite hierarchy of conservation laws to exceptionally constrained particle scattering. In flat space, this connection is well understood. The existence of  infinitely many commuting local higher-spin charges in a massive quantum field theory (QFT) implies the absence of particle production and factorization of the multi-particle $S$-matrix, under standard locality and asymptotic-state assumptions \cite{Zamolodchikov:1977nu,Zamolodchikov:1978xm,Shankar:1977cm,Parke:1980ki}. In turn, imposing no particle production at tree level can determine the interaction potential, providing a constructive principle for identifying integrable theories \cite{Arefeva:1974bk,Braden:1991vz,Dorey:1996gd,Gabai:2018tmm}.

Whether such a constructive principle survives spacetime curvature is  much less clear, with no successful example discovered so far. It is especially interesting to explore this question in the maximally symmetric case of AdS$_2$, particularly in relation to the AdS/CFT correspondence. On the one hand, the charge-based approach was explored as early as \cite{Freedman:1990}. The analysis reveals an important subtlety: covariantly conserved tensors need not furnish additional conserved charges. On the other hand, while AdS boundary correlators provide natural analogues of amplitudes, the conventional asymptotic scattering framework is
unavailable. This makes it difficult to translate the flat-space statement of no particle production into AdS. Nevertheless, one can still ask: is there a condition intrinsic to AdS correlators that plays the constructive role of no particle production?

In this paper, we provide an affirmative answer to this question. We propose that the AdS analogue of the flat-space condition, or more precisely a working definition, is the absence of logarithmic singularities in arbitrary $n$-point functions. In fact, there is a natural motivation for this criterion. Logarithmic singularities in conformal correlators are associated with anomalous dimensions of multi-particle operators, or equivalently with interaction-induced shifts of multi-particle energies in global AdS. Requiring their absence is therefore in line with the highly constrained spectrum expected of an integrable theory.

As a concrete example, we consider a single scalar in AdS${}_2$ with arbitrary mass. We show that imposing this condition diagrammatically at arbitrary multiplicity uniquely fixes the mass $m^2=2$ and reconstructs the Liouville theory. The key ingredient is an all-multiplicity formula for the AdS Berends--Giele currents. It also yields closed-form expressions for general $n$-point boundary correlators, which reproduce the Virasoro stress-tensor correlators at leading order in the large-central-charge expansion. In a complementary approach, we analyze the most general local spin-four current and show that its full covariant conservation independently reproduces the same potential and mass. In particular, the conclusion does not contradict the no-go theorem of \cite{Antunes:2025}.

We emphasize that the logarithm cancellation condition is neither hand-picked nor specific to Liouville: the same diagrammatic mechanism extends to multi-scalar theories and reconstructs the $A_N$ Toda potentials. The higher-spin conservation analysis generalizes in parallel and independently selects the same theories. We outline both extensions here; a comprehensive treatment will appear in the longer companion paper \cite{longpaper}.
 
\section{No particle production in AdS}

To be concrete, let us consider a single scalar $\phi$ in AdS$_2$
\begin{equation}\label{singlephiL}
\mathcal{L}=\frac12(\nabla\phi)^2+W(\phi)\;,
\end{equation}
where the potential has the power series expansion
\begin{equation}\label{Wexpansion}
  W=\frac{1}{2}m^2\phi^2+\sum_{n=3}^\infty\frac{g_n}{n!}\phi^n\;. 
\end{equation}
Here and below we set the AdS radius to one. We further take $m^2=2$, corresponding to $\Delta=2$ for the dual operator. For the moment, this choice can simply be regarded as a technical convenience which considerably simplifies the AdS perturbation theory. Interestingly, we will see below that the same value can also be selected directly from logarithm cancellation at four points, and later independently from the existence of a higher-spin conserved current.

Given (\ref{Wexpansion}), tree-level correlators are organized as a sum of Witten diagrams. In what follows, we will demonstrate how cancellation of logarithmic singularities uniquely fixes the potential (\ref{Wexpansion}) to be that of Liouville theory.

\vspace{0.5cm}
\noindent{\bf Four- and five-point processes.} Let us start with the simplest cases where all Witten diagrams can be computed explicitly. We denote the bulk-to-bulk and bulk-to-boundary propagators respectively by
\begin{equation}
    \begin{split}
        {}&G(Z,Y)=\frac{1}{2\pi}\left[\frac{\chi}{2}\log\frac{\chi+1}{\chi-1}-1\right]\;,\\
        {}&K_i(Z)= k_i^2(Z)=\left(\frac{z}{z^2+(x-x_i)^2}\right)^2\;,
    \end{split}
\end{equation}
where $Z=(z,x)$ and $Y=(z',x')$ are bulk points in Poincar\'e coordinates with metric $ds^2=z^{-2}(dz^2+dx^2)$, and $\chi=(z^2+z'^2+(x-x')^2)/(2zz')$. Define 
\begin{equation}
    L=2-z^2(\partial_z^2+\partial_x^2)\;.
\end{equation}
The propagators then satisfy
\begin{equation}
    LK_i(Z)=0\;,\quad LG(Z,Y)=\delta(Z,Y)\;.
\end{equation}
A useful identity is \cite{DHoker:1999mqo}
\begin{equation}\label{cubicidentity}
   \int_{{\rm AdS}_2} dY G(Z,Y)K_1(Y)K_2(Y)=\frac{k_1(Z)k_2(Z)}{4s_{12}}\;, 
\end{equation}
where $x_{ij}=x_i-x_j$ and $s_{ij}=x_{ij}^2$. At four points, there are exchange Witten diagrams in three channels, proportional to $g_3^2$, together with the  contact Witten diagram which is multiplied by $g_4$. Using (\ref{cubicidentity}), it is clear that the s-channel exchange Witten diagram becomes (and likewise in other channels)
\begin{equation}
    \vTwoDiagram{
      \draw[vTwo internal] (-.33,0)--(.33,0);
      \vTwoLeg{(-.33,0)}{135}{1}\vTwoLeg{(-.33,0)}{225}{2}
      \vTwoLeg{(.33,0)}{315}{3}\vTwoLeg{(.33,0)}{45}{4}
      \node[vTwo bulk] at (-.33,0) {};
      \node[vTwo bulk] at (.33,0) {};
      \node[font=\scriptsize,inner sep=0pt] at (-.33,-.35) {$Y$};
      \node[font=\scriptsize,inner sep=0pt] at (.33,-.35) {$Z$};
    }\xlongequal{\eqref{cubicidentity}}\frac{1}{4s_{12}}\,
    \vTwoDiagram{
      \vTwoAuxLeg{(0,0)}{135}{1}\vTwoAuxLeg{(0,0)}{225}{2}
      \vTwoLeg{(0,0)}{315}{3}\vTwoLeg{(0,0)}{45}{4}
      \node[vTwo bulk] at (0,0) {};
      \node[font=\scriptsize,inner sep=0pt] at (0,-.35) {$Z$};
    }= \frac{1}{4s_{12}}D_{1122}\;.
\end{equation}
Here orange and blue lines denote $k_i$ ($\Delta=1$) and $K_i$ ($\Delta=2$) respectively. After performing the $Y$ integral,  the remaining
$Z$ integral defines the contact $D$-function
\begin{equation}
\nonumber    D_{\Delta_1\ldots \Delta_n}=\int_{{\rm AdS}_{d+1}} dZ \prod_{i=1}^n k_i^{\Delta_i}(Z)\;.
\end{equation}
These $D$-functions satisfy the differential recursion relations
\begin{equation}\label{Dfundiffrecur}
    \frac{\partial}{\partial x_{ij}^2}D_{\Delta_1\ldots \Delta_n}=\frac{2\Delta_i\Delta_j}{d-\sum_{\ell=1}^n \Delta_\ell}D_{\Delta_1\ldots \Delta_i+1\ldots \Delta_j+1\ldots \Delta_n}\;.
\end{equation}
Consequently, both types of Witten diagrams can be reduced to the basic function $D_{1111}$. The latter can be explicitly evaluated as \cite{Usyukina:1992jd} 
\begin{equation}\label{D1111}
\nonumber    D_{1111}=\frac{\pi^{\frac{d}{2}}\Gamma(\frac{4-d}{2})}{2s_{13}s_{24}}\frac{2({\rm Li}_2(\zeta)-{\rm Li}_2(\bar{\zeta}))+\log(\zeta\bar{\zeta})\log\frac{1-\zeta}{1-\bar{\zeta}}}{\zeta-\bar{\zeta}}\;,
\end{equation}
where $\zeta\bar{\zeta}=\frac{s_{12}s_{34}}{s_{13}s_{24}}$ and  $(1-\zeta)(1-\bar{\zeta})=\frac{s_{14}s_{23}}{s_{13}s_{24}}$.
Note that $d$ only enters via the overall prefactor. For $d=1$, the two cross ratios become dependent, with $\zeta=\bar{\zeta}=\eta$, and  the dilogarithms in $D$-functions reduce to  logarithms. Taking into account the combinatoric factors, cancellation of logarithms requires
\begin{equation}\label{eqn:g4val}
    g_4=\frac{1}{2}g_3^2\;.
\end{equation}
The same cancellation also allows us to single out $\Delta=2$. Although the Witten diagrams are no longer elementary for generic $\Delta$, their logarithmic singularities can be computed recursively in the conformal block decomposition \cite{Zhou:2018sfz}. Requiring their cancellation fixes $\Delta=2$, together with (\ref{eqn:g4val}).

\begin{figure}[!ht]
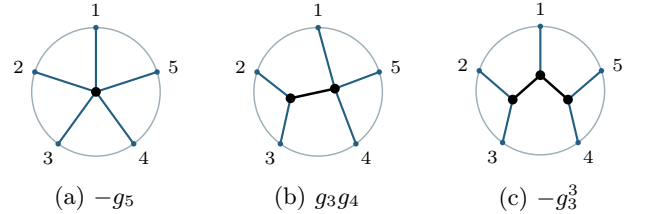

  \centering
  \begin{minipage}{.32\columnwidth}
    \centering
    $\vTwoDiagram[.85]{
      \foreach \angle/\label in {90/1,162/2,234/3,306/4,18/5}
        {\vTwoLeg{(0,0)}{\angle}{\label}}
      \node[vTwo bulk] at (0,0) {};
    }$\par
    \small (a) $-g_5$
  \end{minipage}\hfill
  \begin{minipage}{.32\columnwidth}
    \centering
    $\vTwoDiagram[.85]{
      \draw[vTwo internal] (-.43,-.10)--(.26,.05);
      \vTwoLeg{(-.43,-.10)}{162}{2}\vTwoLeg{(-.43,-.10)}{234}{3}
      \vTwoLeg{(.26,.05)}{90}{1}\vTwoLeg{(.26,.05)}{306}{4}
      \vTwoLeg{(.26,.05)}{18}{5}
      \node[vTwo bulk] at (-.43,-.10) {};
      \node[vTwo bulk] at (.26,.05) {};
    }$\par
    \small (b) $g_3g_4$
  \end{minipage}\hfill
  \begin{minipage}{.32\columnwidth}
    \centering
    $\vTwoDiagram[.85]{
      \draw[vTwo internal] (-.43,-.14)--(0,.24)--(.43,-.14);
      \vTwoLeg{(0,.24)}{90}{1}
      \vTwoLeg{(-.43,-.14)}{162}{2}\vTwoLeg{(-.43,-.14)}{234}{3}
      \vTwoLeg{(.43,-.14)}{306}{4}\vTwoLeg{(.43,-.14)}{18}{5}
      \node[vTwo bulk] at (-.43,-.14) {};
      \node[vTwo bulk] at (0,.24) {};
      \node[vTwo bulk] at (.43,-.14) {};
    }$\par
    \small (c) $-g_3^3$
  \end{minipage}
  \caption{The three five-point tree topologies: (a) quintic contact,
  (b) cubic--quartic exchange, and (c) three cubic vertices. The full
  correlator sums over $1$, $10$, and $15$ inequivalent
  boundary labelings, respectively. Blue lines represent $K_i$, black internal lines
  represent $G$, and filled black dots are integrated bulk vertices.
  The displayed factors give the interaction weights of each labeled tree.}
  \label{fig:five-point-trees}
\end{figure}

The five-point calculation proceeds similarly. There are now three distinct topologies, shown in Fig. \ref{fig:five-point-trees}. Repeatedly using (\ref{cubicidentity}) and (\ref{Dfundiffrecur}), we reduce all Witten diagrams to the basic contact diagram $D_{11112}$, which in turn can be expressed as linear combinations of $D_{1111}$ where four of the five external points are selected at a time \cite{Goncalves:2019znr}. In the 1d limit we again only have logarithms, and their cancellation fixes the quintic coupling to be
\begin{equation}
    g_5=\frac{1}{4}g_3^3\;.
\end{equation}
These two cases suggest the general pattern
\begin{equation}  \label{eqn:gnValLiouville}
    g_n=2^{3-n}g_3^{\,n-2}\;.
\end{equation}
Proving it directly is impractical because the higher-point Witten diagrams quickly become prohibitively complicated. As we now show, however, a simple
all-multiplicity identity allows us to bypass their explicit evaluation.

\vspace{0.5cm}
\noindent{\bf An analytic diagrammatic identity.} Let $A$ be a set of $n$ boundary points and define $\Phi_A(X)$ to be the sum of all tree diagrams connecting the points in $A$ to a fixed bulk point $X$. For the simplest case with $n=1$, we have 
\begin{equation}
    \Phi_{\{i\}}(X)=K_i(X)\;.
\end{equation}
More generally, we have the recursive definition
\begin{equation}\label{PhiALHS}
    \Phi_A(X)=-\int dY G(X,Y)\sum_{r=2}^n g^*_{r+1}\sum_{\pi\in\Pi_r(A)}\prod_{B\in\pi}\Phi_B(Y)\;,
\end{equation}
where $\Pi_r(A)$ denotes the unordered partitions of $A$ into $r$ nonempty blocks. In fact, $\Phi_A(X)$ is precisely the Berends--Giele current. Quite remarkably, for special couplings $g_n^{*}=2^{3-n}g_3^{\,n-2}$
\begin{equation}\label{analytictreeid}
    \Phi_A(X)=\left(-\frac{g_3}{8}\right)^{n-1}\sum_{\sigma\in{\rm Perm}(A)}\frac{k_{\sigma_1}(X)k_{\sigma_n}(X)}{s_{\sigma_1\sigma_2}\ldots s_{\sigma_{n-1}\sigma_n}}\;.
\end{equation}
For example, the immediate generalization of (\ref{cubicidentity}) is $n=3$ where the diagrams are
\begin{equation} \label{eqn:4ptId}
    \Phi_{\{1,2,3\}}=-g_4^*\,
    \vTwoDiagram[.60]{
      \draw[vTwo internal] (-.62,0)--(.12,0);
      \vTwoLeg{(.12,0)}{90}{1}\vTwoLeg{(.12,0)}{0}{2}
      \vTwoLeg{(.12,0)}{270}{3}
      \node[vTwo bulk] at (.12,0) {};
      \vTwoFixedRoot
    }
    +g_3^2\ \sum_{\rm cyc}
    \vTwoDiagram[.60]{
      \draw[vTwo internal] (-.62,0)--(-.12,0)--(.42,-.14);
      \vTwoLeg{(-.12,0)}{100}{1}\vTwoLeg{(.42,-.14)}{15}{2}
      \vTwoLeg{(.42,-.14)}{280}{3}
      \node[vTwo bulk] at (-.12,0) {};
      \node[vTwo bulk] at (.42,-.14) {};
      \vTwoFixedRoot
    }\;.
\end{equation}
The identity gives
\begin{equation} \label{eqn:4ptId-res}
    \Phi_{\{1,2,3\}}=\frac{g_3^2}{32}\left(\frac{k_1k_2}{s_{13}s_{23}}+\frac{k_1k_3}{s_{12}s_{23}}+\frac{k_2k_3}{s_{12}s_{13}}\right)\;.
\end{equation}
This follows directly from \eqref{cubicidentity} and the algebraic identity
\begin{equation}
\begin{split}
\nonumber  K_1K_2K_3={}&\frac{1}{2}\left(\frac{K_1k_2k_3}{s_{23}}+\frac{K_2k_1k_3}{s_{13}}+\frac{K_3k_1k_2}{s_{12}}\right)\\
    {}&-\frac{1}{4}\left(\frac{s_{12}K_1K_2}{s_{13}s_{23}}+\frac{s_{13}K_1K_3}{s_{12}s_{23}}+\frac{s_{23}K_2K_3}{s_{12}s_{13}}\right)\;.
\end{split}
\end{equation}
The proof of (\ref{analytictreeid}) for general $n$ is outlined in Sec.~\ref{apd:proof} of the Supplemental Material. Another important identity follows from acting on (\ref{analytictreeid}) with the equation of motion operator
\begin{equation}\label{analytictreeideom}
    L\Phi_A(X)=\left(-\frac{g_3}{8}\right)^{n-1}\!\!\!\!\!\!\!\!\sum_{\sigma\in{\rm Perm}(A)}\!\!\!\!\!\!\! \frac{4s_{\sigma_1\sigma_n}K_{\sigma_1}(X)K_{\sigma_n}(X)}{s_{\sigma_1\sigma_2}\ldots s_{\sigma_{n-1}\sigma_n}}\;. 
\end{equation}

Two comments are in order. First, the expression (\ref{analytictreeid}) holds for special values $g_n^*$ which turn out to be the Liouville couplings. This, however, does not make the bootstrap argument circular, since at each order all lower-point couplings have already been fixed from lower-point correlators. Second, the Berends--Giele organization (\ref{PhiALHS}) of the AdS diagrams does not by itself imply the remarkable simplification in (\ref{analytictreeid}). The latter results from a special conspiracy among the couplings.

\vspace{0.5cm}
\noindent{\bf All-order interactions via recursion.} We can now bootstrap the interactions to all orders. Let us suppose that all couplings $g_m$ with $m<n$ have been fixed to their special values $g_m^*$, and consider the $(n-1)$-point current $\Phi_{\{1,\ldots,n-1\}}$. After acting with the operator $L$, we glue it to the remaining bulk-to-boundary propagator and call the result $G_n^*$. Using (\ref{analytictreeideom}) together with
\begin{equation}
    D_{222}(x_i,x_j,x_k)=\frac{3\pi}{8s_{ij}s_{jk}s_{ki}},
\end{equation}
we immediately get 
\begin{equation}\label{Gnratexp}
    G_n^*=\frac{3\pi}{2}\left(-\frac{g_3}{8}\right)^{n-2}\sum_{\sigma\in S_{n-1}}\frac{1}{s_{n\sigma_1}s_{\sigma_1\sigma_2}\ldots s_{\sigma_{n-1}n}}\;,
\end{equation}
which is manifestly rational. On the other hand, the exchange contributions to the physical correlator $G_n$ and $G_n^*$ are identical, since they involve only the already fixed lower-point couplings. Their difference is therefore entirely due to the $n$-point contact diagram
\begin{equation}
    G_n-G_n^*=-(g_n-2^{3-n}g_3^{n-2})D_{22\ldots22}\;.
\end{equation}
The contact diagram contains logarithmic singularities for $n>3$ while $G_n^*$ does not. Requiring their absence fixes $g_n$ to the value in \eqref{eqn:gnValLiouville}. 
Iterating this argument determines all couplings and yields the Liouville potential
\begin{equation} \label{eqn:LiouvillePotential}
    W(\phi)=\frac{8}{g_3^2}\left(e^{\frac{g_3\phi}{2}}-1-\frac{g_3\phi}{2}\right)\;.
\end{equation}

We should point out that the  cancellation of logarithms at four points has previously been noticed in \cite{Ouyang:2019xdd}. Here, however, we use the cancellation mechanism as a condition to fix the vertices order by order, and demonstrate the absence of such singularities in all $n$-point functions. In \cite{Ouyang:2019xdd} at tree level and  \cite{Beccaria:2019stp} at one loop, the four-point correlator was recognized as the stress tensor correlator $\langle TTTT\rangle$  of the Virasoro algebra. A semiclassical argument was also given in \cite{Beccaria:2019stp} to support this identification at all multiplicities. Our result (\ref{Gnratexp}) confirms this by direct computation at leading order in the large central charge expansion. Moreover, the identities (\ref{analytictreeid}) and (\ref{analytictreeideom}) are more general than the application above, since the corresponding currents can be glued to other Witten diagrams, leading to further diagrammatic relations.

\section{Higher-spin conserved currents}
We now approach the reconstruction problem independently through local conservation laws. For a single real scalar, we determine which potentials $W(\phi)$ admit a nontrivial fully symmetric rank-four current $\mathcal{J}_{\mu\nu\rho\sigma}$ in $\mathrm{AdS}_2$.

A crucial difference from the flat-space analysis is that, in AdS, conservation of the top component $\mathcal{J}_{++++}$, together with the AdS isometries, requires conservation of the full tensor \cite{Antunes:2025}. In the lightcone frame defined by $x^\pm=z\pm i x$, the conservation equations include
\begin{equation}
\begin{aligned}
\nabla_-\mathcal J_{++++}+\nabla_+\mathcal J_{-+++}&=0,\\
\nabla_-\mathcal J_{-+++}+\nabla_+\mathcal J_{--++}&=0,
\end{aligned}
\label{eqn:full_conservation}
\end{equation}
together with the equations obtained by $+\leftrightarrow-$ \footnote{
The full set of conservation equations is invariant under this exchange,
but the current need not be. We allow terms constructed with the volume
form $\epsilon_{\mu\nu}$, which changes sign under the exchange.
}.
The second equation provides an additional constraint that is absent in the usual flat-space analysis and, as we will see, excludes the sinh-Gordon solution allowed there. 

To solve these conservation equations, we consider the most general local, fully symmetric rank-four current of derivative degree at most four, with arbitrary smooth field-dependent coefficients \footnote{The reason for restricting the ansatz to four derivatives is that higher-derivative terms necessarily involve mixed $+$ and $-$ derivatives.
Using the equations of motion and integration by parts, such terms can be reduced to lower-derivative contributions. The remaining total-derivative terms can then be removed consistently from the current by appropriate improvement terms.}. Tracelessness does not need to be imposed separately, as it follows from the solution. We then quotient out contributions generated by scalar improvements and descendants of the stress tensor. Imposing \eqref{eqn:full_conservation} on the remaining representative constrains the allowed form of the potential $W(\phi)$. Since the computation is standard, we defer the details to the Supplemental Material, Sec.~\ref{apd:currents}.

Reducing the coefficients of all five components leaves two constant triples $\mathfrak c_{i,\sigma}$, one for each
$\sigma=\pm$, satisfying
\begin{equation}
\mathfrak c_{1,\sigma}f''-\mathfrak c_{2,\sigma}f'=0,\quad
\mathfrak c_{2,\sigma}(f'-2)+4\mathfrak c_{3,\sigma}f=0,
\label{eqn:master}
\end{equation}
where $f=W'$. A nontrivial nonlinear current requires at least one nonzero triple. Equations \eqref{eqn:master}, with $W(0)=W'(0)=0$, then uniquely select the Liouville potential \eqref{eqn:LiouvillePotential}, with $g_3=2\mathfrak c_{2,\sigma}/\mathfrak c_{1,\sigma}$ for either nonzero triple \footnote{The separate free branch is $W=\tfrac12m^2\phi^2$, with arbitrary mass.}. In particular, conservation fixes $m^2=W''(0)=2$, and hence the standard boundary dimension $\Delta=2$, independently of the earlier correlator analysis. After subtracting improvement terms and conserved descendants, only the purely chiral components of the current remain nonzero, namely
\begin{equation} 
\begin{aligned}
\mathcal J_{++++} & =- \mathfrak c_{1,+} \left[\nabla_+^2\phi-\frac{g_3}{4}(\nabla_+\phi)^2 \right]^2,\\
\mathcal J_{----} & = - \mathfrak c_{1,-} \left[\nabla_-^2\phi-\frac{g_3}{4}(\nabla_-\phi)^2 \right]^2.
\end{aligned}
\label{eqn:current_representative}
\end{equation}
All mixed components containing both $+$ and $-$ indices vanish. The constants $\mathfrak c_{1,+}$ and $\mathfrak c_{1,-}$ are independent, with at least one of them nonzero. Impose the standard near-boundary condition $\phi|_{z \rightarrow 0} = O(z^2)$ as $z \rightarrow 0$. In the absence of boundary degrees
of freedom, vanishing boundary flux for the conserved charge associated with the Killing tensor $\partial_x \otimes \partial_x \otimes\partial_x$ requires $
\mathfrak c_{1,+}=\mathfrak c_{1,-}.$

The resulting current is proportional to the square of the improved chiral stress tensor, realizing the known higher-spin structure of a bulk CFT on $\mathrm{AdS}_2$ \cite{Bazhanov:1994ft,Antunes:2025}. The nontrivial result here is the converse statement: within the class of scalar-potential theories considered above, imposing full spin-four conservation uniquely selects the Liouville potential and fixes $m^2=2$, without assuming conformal invariance.

This conclusion is consistent with the general no-go theorem of \cite{Antunes:2025}. That theorem shows that, for generic $\Delta$, no interacting deformation of a free scalar field in $\mathrm{AdS}_2$ preserves a higher-spin current. An interacting theory admitting such a current must therefore occur at a special, non-generic value of $\Delta$, a possibility left open in \cite{Antunes:2025}. Within the class of theories \eqref{singlephiL}, we find that Liouville theory at $\Delta=2$ is the unique interacting realization.

\section{Generalization to Toda theories}

We extend the conservation argument to $N\geq2$ real scalar fields $\phi^I$ on $\mathrm{AdS}_2$, with Lagrangian
\begin{equation}
\mathcal{L}^{[N]}=
\frac12\sum_{I=1}^N(\nabla\phi^I)^2+W(\phi^1,\phi^2,\ldots,\phi^N).
\end{equation}
For multiple scalar species, a nontrivial cubic higher-spin current is allowed, making rank three the natural starting point. We therefore consider the most general local, fully symmetric rank-three current $\mathcal{J}^{[N]}_{\mu\nu\rho}$ containing at most three derivatives, with arbitrary smooth field-dependent coefficients, and impose the conservation condition
$
\nabla^\mu\mathcal{J}^{[N]}_{\mu\nu\rho}=0
$
on shell.

The computational strategy is essentially the same as in the previous section. We first write down the most general three-derivative ansatz and then quotient out the most general improvement terms. Imposing current conservation subsequently constrains the allowed form of the potential.

To exclude disconnected solutions, we further require that the Hessian
\begin{equation}
H_{IJ}(\phi^1,\phi^2,\ldots,\phi^N)=\partial_{\phi^I}\partial_{\phi^J} W(\phi^1,\phi^2,\ldots,\phi^N)
\end{equation}
does not admit a nontrivial block decomposition under any constant orthogonal transformation. This condition rules out theories that decompose into independent sectors. For instance, when $N=2$, it excludes a disconnected potential of the form
$
W(\phi^1,\phi^2)=W_1(\phi^1)+W_2(\phi^2).
$
For standard Toda theories, this condition is equivalent to requiring the Dynkin diagram to be connected.

Under this assumption, current conservation fixes the potential to be of $A_N$ Toda form
\footnote{The stable vacuum condition \(\partial_{\phi_I}W|_{\phi=0}=0\) imposes
$
b^2\sum_a \mu_a\alpha_a=2\rho.
$
This condition is also discussed in \cite{Ouyang:2019xdd}.}
\begin{equation} \label{eqn:potentialANtoda}
W(\phi^1,\phi^2,\ldots,\phi^N)
=
\sum_{a=1}^N\mu_a \left(e^{b\alpha_a\cdot\phi} -1 \right)
-\frac{2}{b}\,\rho \cdot\phi,
\end{equation}
where $b\neq0$, $\mu_a\neq0$, and
\begin{equation}
\alpha_a\cdot\alpha_b
=
2\delta_{ab}-\delta_{a,b+1}-\delta_{a+1,b},
\quad
\rho = \frac{1}{2} \sum_{\alpha \in \Delta_+} \alpha.
\end{equation}
The vectors $\alpha_a$ are the simple roots of $A_N$ in the $N$-dimensional field space, while $\rho$ is the corresponding Weyl vector, defined as half the sum over all positive roots of $A_N$ \footnote{The proof can be generalized to other Lie algebras by imposing conservation of their lowest higher-spin currents. For a summary of the spectra of conformal dimensions in Toda field theories, see Table 1 of \cite{Ouyang:2019xdd}.}. Whereas the single-scalar Liouville case considered above contains only a field with \(\Delta=2\), the \(A_N\) Toda theory described by \eqref{eqn:potentialANtoda} contains a larger scalar spectrum with boundary dimensions \(\Delta=2,3,\ldots,N+1\).
It is worth emphasizing that the exponential dependence, the \(A_N\) root geometry, and the curvature-dependent linear term all follow directly from current conservation, without assuming an exponential form for the potential.

\begin{figure}
    \includegraphics[width=\linewidth]{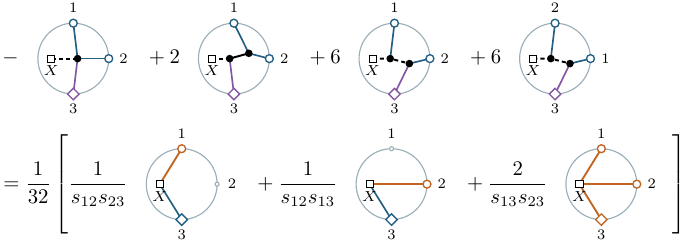}
    \caption{The identity generalizes the single-scalar construction in \eqref{eqn:4ptId}-\eqref{eqn:4ptId-res} to scalar fields with boundary dimensions \(\Delta=2\) and \(\Delta=3\). Here, orange, blue, and purple denote $k_i$, $k_i^2=K_i$, and $k_i^3$, respectively. Circles and diamonds represent the $\Delta=2$ and $\Delta=3$ boundary insertions, respectively. All other conventions are the same as those defined previously.}
    \label{fig:4ptIdGenN2}
\end{figure}

The diagrammatic techniques used in the single-scalar analysis also generalize to the multi-scalar case. For simplicity, in Fig.~\ref{fig:4ptIdGenN2} we present the first nontrivial identity for $N=2$ that involves both scalar fields. Note that the last diagram in the second line leads to higher-point $D$-functions which may seem to spoil logarithm cancellation. However, we can prove that after the $L$ action and the bulk-to-boundary propagator gluing, the $D$-function combination coming from the last diagram reduces to a rational function after using $D$-function identities \footnote{These identities generalize (4.55) of \cite{Rastelli:2017udc}.}. Rationality is generally guaranteed by these identities both for higher-point BG currents and other Toda theories. The general identities for the multi-scalar BG currents, the proof of rationality, and the reconstruction of the
Lagrangians will be presented in detail in the companion paper \cite{longpaper}.

\section{Discussion and outlook}

In this paper, we showed how the absence of logarithmic singularities in correlators determines the Lagrangian of AdS$_2$ theories to all orders of interactions. This is reminiscent of no particle production in flat space and therefore gives rise to one notion of integrability in AdS. However, other notions of integrability are needed to accommodate $\frac{1}{2}$-BPS Wilson loops, which are believed to
be integrable even though their correlators contain uncanceled logarithms \cite{Giombi:2017cqn}. It would therefore be very interesting to clarify or even classify different notions of integrability in AdS$_2$ in future work. 

There are many immediate extensions of the results in this paper. We have demonstrated the cancellation of logarithms in tree-level Liouville correlators. It would be interesting to show this  also happens at general multiplicities for loop corrections, which can be achieved presumably by extending the diagrammatic identities (\ref{analytictreeid}). In addition to the Toda theories to be discussed in \cite{longpaper}, we expect our strategy to extend to Wess--Zumino--Witten models in AdS$_2$ and to supersymmetric theories, whose four-point correlators were studied in \cite{Beccaria:2019dju,Beccaria:2020qtk}.

Although flat space motivated and guided this work, our AdS results, in turn, offer new insights. As we will elaborate in \cite{longpaper}, we also find identities for flat-space Feynman diagrams that are inspired by, and structurally similar to, \eqref{analytictreeid}. These identities allow us to efficiently determine the Lagrangian of integrable theories to all orders from no particle production, without needing to take the multi-Regge limits as in \cite{Gabai:2018tmm,Dorey:2021hub}.

Finally, there is also an amplitude perspective. The Virasoro correlators reproduced by our tree-level AdS$_2$ correlators also arise in the chiral boundary-graviton sector of AdS$_3$, where they admit a BCFW-like formulation \cite{Mathys:2026pdq}. We derive analogous all-multiplicity diagrammatic identities for boundary-anchored dressed Berends--Giele currents in the chiral Chern--Simons and gravitational sectors of AdS$_3$ \cite{AdS3result}. These identities provide a bulk realization of the corresponding recursion and exhibit a simple squaring relation between the spin-one and spin-two responses. They provide a concrete setting for comparing how the same boundary recursive structure emerges from special scalar interactions in AdS$_2$ and gauge symmetry in AdS$_3$.

\vspace{0.3cm}

\noindent{\bf Acknowledgments.} We thank Pietro Ferrero for collaboration at an early stage of this project, and Antonio Antunes, Leonardo Rastelli and Aninda Sinha for helpful discussions. We are also grateful to Yunfeng Jiang and Aninda Sinha 
for valuable comments on the draft. The work of X.Z. is supported by NSFC Grant No. 12275273 and funds from the Chinese Academy of Sciences. This work is also supported by NSFC Grant No. 12247103.

\bibliography{references}

@article{Freedman:1990,
  author       = {Freedman, D. Z. and Lerda, A. and Penati, S.},
  title        = {Covariant Form for the Conserved Currents of the
                  Sine-Gordon and Liouville Theories},
  journal      = {Phys. Lett. B},
  volume       = {250},
  number       = {1--2},
  pages        = {102--106},
  year         = {1990},
  doi          = {10.1016/0370-2693(90)91161-4}
}

@article{Antunes:2025,
  author       = {Antunes, Ant{\'o}nio and Levine, Nat and Meineri, Marco},
  title        = {Demystifying Integrable {QFTs} in {AdS}: No-Go Theorems for
                  Higher-Spin Charges},
  journal      = {SciPost Phys.},
  volume       = {20},
  number       = {3},
  pages        = {088},
  year         = {2026},
  doi          = {10.21468/SciPostPhys.20.3.088},
  eprint       = {2502.06937},
  archivePrefix = {arXiv},
  primaryClass = {hep-th}
}

@article{Goncalves:2019znr,
    author = "Gon{\c{c}}alves, Vasco and Pereira, Raul and Zhou, Xinan",
    title = "{$20'$ Five-Point Function from $AdS_5\times S^5$ Supergravity}",
    eprint = "1906.05305",
    archivePrefix = "arXiv",
    primaryClass = "hep-th",
    reportNumber = "PUPT-2588",
    doi = "10.1007/JHEP10(2019)247",
    journal = "JHEP",
    volume = "10",
    pages = "247",
    year = "2019"
}

@article{DHoker:1999mqo,
    author = "D'Hoker, Eric and Freedman, Daniel Z. and Rastelli, Leonardo",
    title = "{AdS / CFT four point functions: How to succeed at z integrals without really trying}",
    eprint = "hep-th/9905049",
    archivePrefix = "arXiv",
    reportNumber = "MIT-CTP-2857, UCLA-99-TEP-16",
    doi = "10.1016/S0550-3213(99)00526-X",
    journal = "Nucl. Phys. B",
    volume = "562",
    pages = "395--411",
    year = "1999"
}

@article{Usyukina:1992jd,
	author = {Usyukina, N. I. and Davydychev, Andrei I.},
	doi = {10.1016/0370-2693(93)91834-A},
	journal = {Phys. Lett. B},
	pages = {363--370},
	reportnumber = {NPI-MSU-92-38-287},
	title = {{An Approach to the evaluation of three and four point ladder diagrams}},
	volume = {298},
	year = {1993}}

@article{Ouyang:2019xdd,
    author = "Ouyang, Hao",
    title = "{Holographic four-point functions in Toda field theories in AdS$_{2}$}",
    eprint = "1902.10536",
    archivePrefix = "arXiv",
    primaryClass = "hep-th",
    reportNumber = "NORDITA-2019-015",
    doi = "10.1007/JHEP04(2019)159",
    journal = "JHEP",
    volume = "04",
    pages = "159",
    year = "2019"
}

@article{Beccaria:2019stp,
    author = "Beccaria, Matteo and Tseytlin, Arkady A.",
    title = "{On boundary correlators in Liouville theory on AdS$_{2}$}",
    eprint = "1904.12753",
    archivePrefix = "arXiv",
    primaryClass = "hep-th",
    reportNumber = "Imperial-TP-AT-2019-02",
    doi = "10.1007/JHEP07(2019)008",
    journal = "JHEP",
    volume = "07",
    pages = "008",
    year = "2019"
}

@article{Gabai:2018tmm,
    author = "Gabai, Barak and Maz{\'a}{\v{c}}, Dalimil and Shieber, Andrei and Vieira, Pedro and Zhou, Yehao",
    title = "{No Particle Production in Two Dimensions: Recursion Relations and Multi-Regge Limit}",
    eprint = "1803.03578",
    archivePrefix = "arXiv",
    primaryClass = "hep-th",
    doi = "10.1007/JHEP02(2019)094",
    journal = "JHEP",
    volume = "02",
    pages = "094",
    year = "2019"
}

@article{Giombi:2017cqn,
    author = "Giombi, Simone and Roiban, Radu and Tseytlin, Arkady A.",
    title = "{Half-BPS Wilson loop and AdS$_2$/CFT$_1$}",
    eprint = "1706.00756",
    archivePrefix = "arXiv",
    primaryClass = "hep-th",
    reportNumber = "IMPERIAL-TP-AT-2017-06, PUPT-2526, Imperial-TP-AT-2017-06",
    doi = "10.1016/j.nuclphysb.2017.07.004",
    journal = "Nucl. Phys. B",
    volume = "922",
    pages = "499--527",
    year = "2017"
}

@article{Mathys:2026pdq,
    author = "Mathys, Gr{\'e}goire and Pimentel, Guilherme L. and Rost, Facundo",
    title = "{${{\rm AdS}_3}$ Recursion Relations, Double Copy and ${{\rm CFT}_2}$ Ward Identities}",
    eprint = "2609.12046",
    archivePrefix = "arXiv",
    primaryClass = "hep-th",
    month = "9",
    year = "2026",
    journal = {}
}

@article{Beccaria:2019dju,
    author = "Beccaria, Matteo and Jiang, Hongliang and Tseytlin, Arkady A.",
    title = "{Supersymmetric Liouville theory in AdS$_{2}$ and AdS/CFT}",
    eprint = "1909.10255",
    archivePrefix = "arXiv",
    primaryClass = "hep-th",
    reportNumber = "Imperial-TP-AT-2019-07",
    doi = "10.1007/JHEP11(2019)051",
    journal = "JHEP",
    volume = "11",
    pages = "051",
    year = "2019"
}

@article{Beccaria:2020qtk,
    author = "Beccaria, Matteo and Jiang, Hongliang and Tseytlin, Arkady A.",
    title = "{Boundary correlators in WZW model on AdS$_{2}$}",
    eprint = "2001.11269",
    archivePrefix = "arXiv",
    primaryClass = "hep-th",
    reportNumber = "Imperial-TP-AT-2020-01",
    doi = "10.1007/JHEP05(2020)099",
    journal = "JHEP",
    volume = "05",
    pages = "099",
    year = "2020"
}

@article{Zamolodchikov:1977nu,
    author = "Zamolodchikov, Alexander B. and Zamolodchikov, Alexei B.",
    title = "{Relativistic Factorized S Matrix in Two-Dimensions Having O(N) Isotopic Symmetry}",
    reportNumber = "JINR-E2-10857",
    doi = "10.1016/0550-3213(78)90239-0",
    journal = "JETP Lett.",
    volume = "26",
    pages = "457",
    year = "1977"
}

@article{Zamolodchikov:1978xm,
    author = "Zamolodchikov, Alexander B. and Zamolodchikov, Alexei B.",
    editor = "Khalatnikov, I. M. and Mineev, V. P.",
    title = "{Factorized s Matrices in Two-Dimensions as the Exact Solutions of Certain Relativistic Quantum Field Models}",
    reportNumber = "ITEP-35-1978",
    doi = "10.1016/0003-4916(79)90391-9",
    journal = "Annals Phys.",
    volume = "120",
    pages = "253--291",
    year = "1979"
}

@article{Shankar:1977cm,
    author = "Shankar, R. and Witten, Edward",
    title = "{The S Matrix of the Supersymmetric Nonlinear Sigma Model}",
    reportNumber = "HUTP-77/A076",
    doi = "10.1103/PhysRevD.17.2134",
    journal = "Phys. Rev. D",
    volume = "17",
    pages = "2134",
    year = "1978"
}

@article{Parke:1980ki,
    author = "Parke, Stephen J.",
    title = "{Absence of Particle Production and Factorization of the $S$ Matrix in (1+1)-dimensional Models}",
    reportNumber = "HUTP-80/A021",
    doi = "10.1016/0550-3213(80)90196-0",
    journal = "Nucl. Phys. B",
    volume = "174",
    pages = "166--182",
    year = "1980"
}

@article{Bazhanov:1994ft,
    author = "Bazhanov, Vladimir V. and Lukyanov, Sergei L. and Zamolodchikov, Alexander B.",
    title = "{Integrable structure of conformal field theory, quantum KdV theory and thermodynamic Bethe ansatz}",
    eprint = "hep-th/9412229",
    archivePrefix = "arXiv",
    reportNumber = "CLNS-94-1316, RU-94-98",
    doi = "10.1007/BF02101898",
    journal = "Commun. Math. Phys.",
    volume = "177",
    pages = "381--398",
    year = "1996"
}

@article{Rastelli:2017udc,
    author = "Rastelli, Leonardo and Zhou, Xinan",
    title = "{How to Succeed at Holographic Correlators Without Really Trying}",
    eprint = "1710.05923",
    archivePrefix = "arXiv",
    primaryClass = "hep-th",
    reportNumber = "YITP-SB-2017-44",
    doi = "10.1007/JHEP04(2018)014",
    journal = "JHEP",
    volume = "04",
    pages = "014",
    year = "2018"
}

@article{Arefeva:1974bk,
    author = "Arefeva, Irina and Korepin, Vladimir",
    title = "{Scattering in two-dimensional model with Lagrangian (1/gamma) ((d(mu)u)**2/2 + m**2 cos(u-1))}",
    journal = "Pisma Zh. Eksp. Teor. Fiz.",
    volume = "20",
    pages = "680",
    year = "1974"
}

@inproceedings{Dorey:1996gd,
    author = "Dorey, P.",
    title = "{Exact S matrices}",
    booktitle = "{Eotvos Summer School in Physics: Conformal Field Theories and Integrable Models}",
    eprint = "hep-th/9810026",
    archivePrefix = "arXiv",
    reportNumber = "DTP-98-69",
    pages = "85--125",
    month = "8",
    year = "1996"
}

@article{Braden:1991vz,
    author = "Braden, H. W. and Sasaki, R.",
    title = "{Affine Toda perturbation theory}",
    reportNumber = "YITP-U-91-40, EDINBURGH-90-91-02",
    doi = "10.1016/0550-3213(92)90601-7",
    journal = "Nucl. Phys. B",
    volume = "379",
    pages = "377--428",
    year = "1992"
}

@misc{longpaper,
    author = "Zhong, De-liang and Zhou, Xinan",
    note = "{Work to appear}"
}

@misc{AdS3result,
    author = "Zhong, De-liang and {Zhou}, Xinan",
    note = "{Work in progress}"
}

@article{Dorey:2021hub,
    author = "Dorey, Patrick and Polvara, Davide",
    title = "{Tree level integrability in 2d quantum field theories and affine Toda models}",
    eprint = "2111.02210",
    archivePrefix = "arXiv",
    primaryClass = "hep-th",
    reportNumber = "SAGEX-21-35-E",
    doi = "10.1007/JHEP02(2022)199",
    journal = "JHEP",
    volume = "02",
    pages = "199",
    year = "2022"
}

@article{Zhou:2018sfz,
    author = "Zhou, Xinan",
    title = "{Recursion Relations in Witten Diagrams and Conformal Partial Waves}",
    eprint = "1812.01006",
    archivePrefix = "arXiv",
    primaryClass = "hep-th",
    reportNumber = "PUPT-2575",
    doi = "10.1007/JHEP05(2019)006",
    journal = "JHEP",
    volume = "05",
    pages = "006",
    year = "2019"
}
\clearpage
\onecolumngrid
\begin{center}
\large\bfseries Supplemental Material
\end{center}
\setcounter{section}{0}
\setcounter{subsection}{0}
\setcounter{equation}{0}
\setcounter{figure}{0}
\setcounter{table}{0}
\renewcommand{\thesection}{S\arabic{section}}
\renewcommand{\thesubsection}{\thesection.\arabic{subsection}}
\renewcommand{\theequation}{S\arabic{equation}}
\renewcommand{\thefigure}{S\arabic{figure}}
\renewcommand{\thetable}{S\arabic{table}}
\renewcommand{\theHsection}{S\arabic{section}}
\renewcommand{\theHsubsection}{\theHsection.\arabic{subsection}}
\renewcommand{\theHequation}{S\arabic{equation}}
\renewcommand{\theHfigure}{S\arabic{figure}}
\renewcommand{\theHtable}{S\arabic{table}}
\setcounter{secnumdepth}{2}
\allowdisplaybreaks[2]

\section{Spin-four current conservation in AdS${}_{2}$ and the Liouville potential}
\label{apd:currents}

We explain how full tensor conservation leads to the potential constraints \eqref{eqn:master} and then compute the boundary flux associated with the resulting conserved charge.

\paragraph{Conventions and full ansatz.}

We work in lightcone coordinates on AdS\({}_2\),
$
x^\pm=z\pm i x,\ ds^2=\frac{dx^+dx^-}{z^2},
$
and express all tensors in the normalized null coframe
$
e^\pm=\frac{dx^\pm}{\sqrt{2}z}\, .
$
Define \(p_n=\nabla_+^n\phi\) and \(q_n=\nabla_-^n\phi\). Commuting covariant derivatives gives
\begin{equation}
\nabla_-p_n=\frac12\left[\nabla_+^{n-1}f-n(n-1)p_{n-1}\right],
\qquad \text{where}\qquad
f = W'(\phi)\, .
\label{cur:mixed}
\end{equation}
The corresponding relations for \(q_n\) follow by exchanging \(+\leftrightarrow-\). Primes denote derivatives with respect to \(\phi\).

The fully symmetric rank-four tensor \(\mathcal{J}_{\mu\nu\rho\sigma}\) has five independent components, which we denote by
$$
A_+=\mathcal J_{++++},\qquad
B_+=\mathcal J_{-+++},\qquad
D=\mathcal J_{--++},\qquad
B_-=\mathcal J_{+---},\qquad
A_-=\mathcal J_{----}.
$$
In particular, the ansatz allows tensor structures involving \(\epsilon_{\mu\nu}\) in \(\mathcal{J}_{\mu\nu\rho\sigma}\). Denoting \(z_{+,n}=p_n\), \(z_{-,n}=q_n\), with \(\sigma=\pm\), the most general local polynomial ansatz up to derivative degree four is
\begin{align}
A_\sigma&=A_{\sigma1}z_{\sigma,4}+A_{\sigma2}z_{\sigma,3}z_{\sigma,1}+A_{\sigma3}z_{\sigma,2}^2+A_{\sigma4}z_{\sigma,2}z_{\sigma,1}^2+A_{\sigma5}z_{\sigma,1}^4,\nonumber\\
 B_\sigma&=B_{\sigma1}z_{\sigma,2}+B_{\sigma2}z_{\sigma,1}^2+z_{-\sigma,1}\bigl(B_{\sigma3}z_{\sigma,3}+B_{\sigma4}z_{\sigma,2}z_{\sigma,1}+B_{\sigma5}z_{\sigma,1}^3\bigr),\nonumber\\
 D&=d_0+d_1p_1q_1+d_2p_2q_2+d_3p_2q_1^2+d_4p_1^2q_2+d_5p_1^2q_1^2.
 \label{cur:basis}
\end{align}
All 26 coefficients are initially arbitrary smooth functions of \(\phi\). The reason for considering only the four-derivative case will be explained at the end of this section.

\paragraph{Joint coefficient reduction.}

In components, the conservation equation $\nabla^\mu \mathcal{J}_{\mu\nu\rho\sigma}=0$ becomes
\begin{equation}
 \nabla_-A_++\nabla_+B_+=0,\qquad \nabla_-B_++\nabla_+D=0,\qquad
 \nabla_-D+\nabla_+B_-=0,\qquad \nabla_-B_-+\nabla_+A_-=0.
 \label{cur:divergences}
\end{equation}
Expanding these equations and first matching the five-derivative terms gives a single shared function \(h(\phi)\) and constants \(a_{\sigma i}\):
\begin{align}
 (A_{\sigma1},A_{\sigma2},A_{\sigma3},A_{\sigma4},A_{\sigma5})
 &=(h+a_{\sigma1},4h'+a_{\sigma2},3h'+a_{\sigma3},6h''+a_{\sigma4},h'''+a_{\sigma5}),\nonumber\\
 (B_{\sigma3},B_{\sigma4},B_{\sigma5})&=(-h',-3h'',-h'''),\qquad
 (d_2,d_3,d_4,d_5)=(h',h'',h'',h''').
 \label{cur:triangular}
\end{align}
A common shift of $a_{+1}$ and $a_{-1}$ has been absorbed into \(h\). 

For a symmetric rank-four tensor, the most general scalar improvement \(\mathcal K_{\mu\nu\rho\sigma}\) that is manifestly conserved,
$\nabla^\mu \mathcal K_{\mu\nu\rho\sigma}=0$, and generated by a scalar function \(\Omega(\phi)\), takes the form \cite{Antunes:2025}
\begin{equation}
 \mathcal K_{\mu\nu\rho\sigma}[\Omega]=\nabla_{(\mu}\nabla_\nu\nabla_\rho\nabla_{\sigma)}\Omega
 +g_{(\mu\nu}\nabla_\rho\nabla_{\sigma)}(14-2\Box)\Omega
 +g_{(\mu\nu}g_{\rho\sigma)}(\Box^2-12\Box+9)\Omega.
 \label{cur:improvement}
\end{equation}
Choosing \(\Omega'=h\) removes all \(\phi\)-dependence from the coefficients \(A_{\sigma i}\). Henceforth, we set \(A_{\sigma i}=a_{\sigma i}\), \(B_\sigma=u_\sigma z_{\sigma,2}+v_\sigma z_{\sigma,1}^2\), and \(D=d_0+d_1p_1q_1\). The remaining conservation equations at lower derivative degree then determine
\begin{align}
 u_\sigma&=6a_{\sigma1}-\tfrac12(a_{\sigma1}f'+a_{\sigma2}f),&
 v_\sigma&=\tfrac12[-a_{\sigma1}f''+3a_{\sigma2}-a_{\sigma3}(f'-2)-a_{\sigma4}f],\nonumber\\
 d_1&=\tfrac12(a_{\sigma1}f''+a_{\sigma2}f'),&
 d_0'&=-\tfrac12u_\sigma(f'-2)-v_\sigma f-\tfrac12d_1f.
 \label{cur:central}
\end{align}
Crucially, the same \(d_1\) and \(d_0'\) must satisfy the equations for both \(\sigma=+\) and \(\sigma=-\). A constant scalar improvement removes the integration constant in \(d_0\). Defining
$
(\mathfrak c_{1,\sigma},\mathfrak c_{2,\sigma},\mathfrak c_{3,\sigma})
=(a_{\sigma2}-a_{\sigma3},a_{\sigma4},a_{\sigma5}),
$
and substituting the coefficient relations above, the two conservation equations for each \(\sigma=\pm\) reduce to
\begin{align}
 \nabla_{-\sigma}A_\sigma+\nabla_\sigma B_\sigma
 &=\tfrac12(\mathfrak c_{1,\sigma}f''-2\mathfrak c_{2,\sigma}+4\mathfrak c_{3,\sigma}f)z_{\sigma,1}^3, \label{cur:residuals01} \\ 
 \nabla_{-\sigma}B_\sigma+\nabla_\sigma D
 &=\tfrac12(\mathfrak c_{1,\sigma}f''-\mathfrak c_{2,\sigma}f')z_{\sigma,1}^2z_{-\sigma,1}.
 \label{cur:residuals02}
\end{align}
Setting the coefficients of these independent monomials to zero gives \eqref{eqn:master}. 

The first equation, \eqref{cur:residuals01}, is the standard flat-space conservation condition, which admits both the Liouville and sinh-Gordon potentials. The second equation, \eqref{cur:residuals02}, is the additional condition imposed by full tensor conservation in AdS and excludes the sinh-Gordon solution. 

\paragraph{Potential and reduced current.}

We can now solve the differential constraints \eqref{eqn:master}. For nonlinear \(f\), a nonzero triple has \(\mathfrak c_{1,\sigma}\mathfrak c_{2,\sigma}\ne0\). With \(g_3=2\mathfrak c_{2,\sigma}/\mathfrak c_{1,\sigma}\), the first constraint integrates to \(f'=\kappa+\tfrac{1}{2} g_3 f\), and the second fixes \(\kappa=2\) and \(\mathfrak c_{3,\sigma}=-g_3^2 \mathfrak c_{1,\sigma}/16\). The vacuum conditions \(W(0)=W'(0)=0\) therefore give \eqref{eqn:LiouvillePotential}.

Although the potential is already fixed, the current still contains free parameters. Recall that we start with ten constants \(a_{\sigma i}\). The coefficients \(a_{\sigma1}\) do not enter the potential constraints, while \(a_{\sigma2}\) and \(a_{\sigma3}\) appear only through the combination \(a_{\sigma2}-a_{\sigma3}\). Inspecting the resulting components of \(\mathcal{J}\), we find that the current depends on only five independent parameters
$
a_{+1},\ a_{-1},\
a_0,\ \mathfrak{c}_{1 \pm}
$ where $a_0 \equiv a_{+2}+\tfrac12 g_3 a_{+1}
 =a_{-2}+\tfrac12 g_3 a_{-1}$.

The first three parameters can also be removed. The reason is that, given any conserved symmetric rank-two tensor \(R_{\mu\nu}\), one can construct the following conserved rank-four tensor:
\begin{equation}
\mathcal S_{\mu\nu\rho\sigma}[R]=\frac12\left[\nabla_{(\mu}\nabla_\nu R_{\rho\sigma)}-g_{(\mu\nu}(\Box-9)R_{\rho\sigma)}\right].
\label{cur}
\end{equation}
For the theory at hand, there are three conserved rank-two tensors. The first is the stress tensor,
$
T_{\mu\nu}
=
\nabla_\mu\phi\nabla_\nu\phi
-g_{\mu\nu}\left[\tfrac12(\nabla\phi)^2+W\right].
$
The other two, denoted by \(T^{[\pm]}_{\mu \nu}\), are constructed from the chiral quantities
$$
\chi_\pm
=
z_{\pm,2}-\frac{g_3}{4} z_{\pm,1}^2, \quad \text{where} \quad \nabla_-\chi_+=\nabla_+\chi_-=0 \, .
$$
In particular, \(T^{[+]}\) has only the nonvanishing component
$
T^{[+]}_{++}=\chi_+,
$
while \(T^{[-]}\) is defined analogously by
$
T^{[-]}_{--}=\chi_-.
$
The full current can then be decomposed as
\begin{equation}
 \mathcal J=\mathcal K[\Omega]+a_0 \mathcal S[T]+2a_{+1}\mathcal S[T^{[+]}]+2a_{-1}\mathcal S[T^{[-]}]+\mathcal J_{\mathrm{red}}.
 \label{cur:decomposition}
\end{equation}
The reduced current \(\mathcal J_{\mathrm{red}}\) contains only the purely chiral components and is given in \eqref{eqn:current_representative}. Its tracelessness therefore follows immediately from the absence of mixed \(+\) and \(-\) components.

\paragraph{Boundary charge.}

Once we have the conserved tensor current, contracting it with a rank-three Killing tensor $\mathsf{K}_{\mu \nu \rho}$ gives a conserved vector current. Integrating this vector over a hypersurface \(\Sigma\) defines the charge:
\begin{equation}
 j_\mu[\mathsf{K}]=\mathcal J_{\mu\nu\rho\sigma}\mathsf{K}^{\nu\rho\sigma},\qquad
 \nabla^\mu j_\mu[\mathsf{K}]=0,\qquad Q_{\mathsf{K}}[\Sigma]=\int_\Sigma d\Sigma^\mu\,j_\mu[\mathsf{K}].
 \label{cur:charge-definition}
\end{equation}
For definiteness, choose \(\mathsf{K}=(\partial_x)^{\otimes3}\) and let \(\Sigma_x\) be a constant-\(x\) slice, oriented toward increasing \(x\). In the Poincar\'e coordinates, \(d\Sigma^\mu=\delta_x^\mu dz\), so the reduced current \eqref{eqn:current_representative} gives
\begin{equation}
 Q_{\mathsf{K}}[\Sigma_x]=\int_0^\infty dz\,j_x[\mathsf{K}]\, .
 \label{cur:charge-integral}
\end{equation}
Conservation of the charge requires the boundary contribution to \(\frac{dQ_{\mathsf{K}}}{dx}\) to vanish. Using the boundary asymptotics \(\phi=z^2a(x)+O(z^4)\), we find
\begin{equation}
 0= \frac{dQ_{\mathsf{K}}}{dx}=\lim_{z\to0} j_z[\mathsf{K}]= \frac{9i}{4}(\mathfrak{c}_{1,+}-\mathfrak{c}_{1,-})a(x)^2\, .
 \label{cur:boundary}
\end{equation}
This condition therefore requires
$
\mathfrak{c}_{1,+}=\mathfrak{c}_{1,-}.
$

\paragraph{Finite-order reduction.}

We now justify why it is sufficient to consider currents of derivative degree at most four. Terms with more than four derivatives necessarily involve mixed \(+\) and \(-\) derivatives, whose derivative degree can be reduced using the equations of motion and integration by parts (IBP); see, for example, \eqref{cur:mixed}. The only subtlety is that, for a tensor current, the resulting total-derivative terms cannot be discarded component by component without spoiling conservation. For example, in flat space the conservation equation
$
\partial_-A_++\partial_+B_+=0
$
is preserved under the simultaneous redefinition
$
A_+\rightarrow A_++\partial_+\Theta,
B_+\rightarrow B_+-\partial_-\Theta.
$

The AdS argument follows the same logic, with ordinary derivatives replaced by covariant derivatives and the improvement implemented consistently across all five components of the rank-four tensor. Starting from the highest-derivative terms, conservation forces them to assemble into such an improvement. Removing it lowers the maximal derivative degree, and repeating the procedure eventually reduces the current to the degree-four ansatz \eqref{cur:basis}. We defer the technical details to \cite{longpaper}.

\section{Proof of the all-multiplicity identity} \label{apd:proof}

Without loss of generality, let us set $g_3=1$ for convenience in the Liouville theory identity. Then (\ref{PhiALHS}) becomes
\begin{equation}\label{PhiASM}
    \Phi_A(X)=-\int dY G(X,Y)\sum_{r=2}^n 2^{2-r}\sum_{\pi\in\Pi_r(A)}\prod_{B\in\pi}\Phi_B(Y)\;.
\end{equation}
It is convenient to associate to each label $i\in A$ a commuting nilpotent variable $\epsilon_i$, satisfying $\epsilon_i^2=0$, and to define $\epsilon_B=\prod_{i\in B}\epsilon_i$ for $B\subseteq A$. We can then package the currents as 
\begin{equation}
    \Psi_A=\sum_{\varnothing\neq B\subseteq A}\epsilon_B\Phi_B\;.
\end{equation}
Applying $L$ to (\ref{PhiASM}) gives, coefficient by coefficient
\begin{equation}\label{LPsiA}
    \mathcal{E}(\Psi_A)\equiv L\Psi_A+N(\Psi_A)=0\;,\quad\quad N(\Psi)=4\left(e^{\frac{\Psi}{2}}-1-\frac{\Psi}{2}\right)\;.
\end{equation}
Conversely, solving (\ref{LPsiA}) with the bulk-to-bulk propagator $G$ reproduces (\ref{PhiASM}). Let us now consider the response to adding a new source at $x_m$ to $A$. This can be captured by using a new commuting generator $\eta$ with $\eta^2=0$, and we have 
\begin{equation}
    \Psi_{A\cup\{m\}}=\Psi_A+\eta \psi_m\;,
\end{equation}
with
\begin{equation}
    \psi_m=K_m+\sum_{\varnothing\neq B\subseteq A}\epsilon_B\Phi_{B\cup\{m\}}\;.
\end{equation}
Substituting $\Psi_{A\cup\{m\}}$ in (\ref{LPsiA}) and extracting the $\eta$ coefficient gives
\begin{equation}
    \left(L+N'(\Psi_A)\right)\psi_m=0\;,
\end{equation}
where $N'(\Psi)=\partial_\Psi N(\Psi)$. We claim that $\psi_m$ can be constructed from the packaged lower-point currents using a ``creation'' differential operator
\begin{equation}
    \psi_m^*=K_m+\widehat{D}_m\Psi_A\;,
\end{equation}
where we denote the candidate by $\psi_m^*$. The differential operator is given by 
\begin{equation}
    \widehat{D}_m=\frac{1}{8}\left[\frac{(x-x_m)\partial_x-z\partial_z}{z^2+(x-x_m)^2}+\sum_{i\in A}\left(\frac{1}{x_i-x_m}\partial_{x_i}-\frac{2E_i}{s_{im}}\right)\right]\;.
\end{equation}
Here $E_i$ is a label counting operator: $E_i\epsilon_B=\epsilon_B$ if $i\in B$, and $E_i\epsilon_B=0$ otherwise. It is straightforward to verify the following identities
\begin{equation}
[L,\widehat{D}_m]=\frac{K_m}{2}(L-2)\;,\quad\quad \widehat{D}_m N(\Psi)=N'(\Psi)\widehat{D}_m\Psi\;,\quad\quad N'(\Psi)=\frac{1}{2}N(\Psi)+\Psi\;,
\end{equation}
where the middle equality is the ordinary chain rule and  the last follows from the Liouville form of $N$ in (\ref{LPsiA}). Using these identities, we find
\begin{equation}
    \left(L+N'(\Psi_A)\right)\left(K_m+\widehat{D}_m\Psi_A\right)=\left(\widehat{D}_m+\frac{K_m}{2}\right)\mathcal{E}(\Psi_A)\;,
\end{equation}
where the RHS vanishes by (\ref{LPsiA}). This shows that both $\psi_m$ and $\psi_m^*$ obey the same differential equation. 

To establish $\psi_m=\psi_m^*$, we must also exclude the possibility that their difference is a nontrivial homogeneous solution. This can be proven recursively using boundedness. We assume for smaller sets $B$ that $\Phi_B$ is bounded when $|B|\geq 2$, and that all $\Phi_B$ are rational and translation invariant. Near a boundary point which is not one of their insertions, we write $\Phi_B=z^2 F$, with $F$ smooth. At infinity, we assume $\Phi_B=O(R^{-2})$ as $R\to\infty$, where $R^2=x^2+z^2$. Denoting 
\begin{equation}
    D_m^A=\widehat{D}_m\big|_{E_i=1\;\;(i\in A)}\;,
\end{equation}
the base case is $D^{\{i\}}_mK_i=-k_ik_m/(4s_{im})$, which is bounded because $0<k_ik_m\leq s_{im}^{-1}$. The possible limits are: 
\begin{enumerate}
    \item At a boundary point different from all $x_i$, $i\in A$, and from $x_m$, regularity follows from $D_m^A(z^2F)=O(z^2)$. 
    \item At the new insertion $x_m$, with $a=x-xm$, $d_m=a^2+z^2$, the only apparently singular term is 
    \begin{equation}
        \frac{1}{8}\frac{a\partial_x-z\partial_z}{d_m}(z^2F)=-\frac{z^2F}{4d_m}+\frac{z^2(a\partial_xF-z\partial_zF)}{8d_m}\;,
    \end{equation}
    which is bounded since $0\leq z^2/d_m\leq 1$.
    \item At an old insertion $x_i$, we denote $y=x-x_i$, $r_i^2=y^2+z^2$, and $b_i=x_i-x_m\neq 0$. The dangerous leading term is $(8b_i)^{-1}(\partial_x+\partial_{x_i})$. We can use translation invariance
    \begin{equation}
        \left(\partial_x+\sum_{j\in A}\partial_{x_j}\right)\Phi_A=0\;,
    \end{equation}
    to rewrite the derivative part of $D_m^A$ as
    \begin{equation}
        O(r_i)\partial_x+O(r_i)\partial_z+\frac{1}{8}\sum_{j\in A\setminus\{i\}}\left(\frac{1}{x_j-x_m}-\frac{1}{b_i}\right)\partial_{x_j}\;.
    \end{equation}
    The coefficients of $\partial_x$ and $\partial_z$ vanish linearly with $r_i$. Since the only denominators of $\Phi_A$ that can vanish in this limit are powers of $y^2+z^2$, boundedness implies that its $x$- and $z$-derivatives can grow at most as $1/r_i$. Derivatives with respect to the other $x_j$'s do not change this local counting, and the remaining terms in $D_m^A$ are finite. Hence $D_m^A\Phi_A$ is bounded.
    \item At infinity, the bulk derivatives decay one power of $R$ faster, while the other terms in $D_m^A$ preserve the $O(R^{-2})$ behavior. Therefore, we have $D_m^A\Phi_A=O(R^{-2})$.
\end{enumerate}
The BG current (\ref{PhiASM}) is bounded independently. Near $x_i$, exactly one block in (\ref{PhiASM}) contains $i$. It is either a bounded multi-label current, or $K_i=O(r_i^{-2})$. In the latter case another block must omit $i$ and is $O(z^2)=O(r_i^2)$. This implies that the source is globally bounded. Its Green transform is also bounded because $G(X,Y)>0$ and $\int dY G(X,Y)=1/2$. 

Let $U=\psi_m-\psi_m^*=\sum_{B\subseteq A} \epsilon_BU_B$. We have $U_{\varnothing}=0$. Since $N'(\Psi_A)=\sum_{\varnothing\neq C\subseteq A}\epsilon_CQ_C$, it follows that the coefficient equations are triangular
\begin{equation}
    LU_B=-\sum_{\varnothing\neq C\subseteq B}Q_CU_{B\setminus C}\;.
\end{equation}
Induction on $|B|$ reduces every equation to $LU_B=0$. The usual maximum-principle argument, applied on successively larger regions of Euclidean AdS, shows that an everywhere bounded solution of this free massive equation must vanish. This proves that $\psi_m^*$ and $\psi_m$ are identical, and 
\begin{equation}
    \Phi_{A\cup\{m\}}=D^A_m \Phi_A\;.
\end{equation}
This equation now proves (\ref{analytictreeid}) by induction. When $D_m^A$ acts on a term associated with a particular ordering of the labels in $A$, the resulting terms insert $m$ either at one of the two ends or between two neighboring labels, each with the common factor $-1/8$. Summing over the original orderings and all possible insertion positions produces every ordering of $A\cup\{m\}$ exactly once. Together with $\Phi_{\{i\}}=K_i$, this completes the induction.
\end{document}